# A Low-Cost Multi-Agent System for Physical Security in Smart Buildings

Tiago Fonseca*[1][0000-0002-5592-3107], Tiago Dias[12][0000-0002-1693-7872], João Vitorino[12][0000-0002-4968-3653], Luís Lino Ferreira[1][0000-0002-5976-8853] and Isabel Praça[12][0000-0002-2519-9859]

[1] School of Engineering, Polytechnic of Porto (ISEP/IPP), 4249-015 Porto, Portugal
[2] Research Group on Intelligent Engineering and Computing for Advanced Innovation and Development (GECAD), 4249-015 Porto, Portugal
`{calof,tiada,jpmvo,llf,icp}@isep.ipp.pt`

**Abstract.** Modern organizations face numerous physical security threats, from fire hazards to more intricate concerns regarding surveillance and unauthorized personnel. Conventional standalone fire and intrusion detection solutions must be installed and maintained independently, which leads to high capital and operational costs. Nonetheless, due to recent developments in smart sensors, computer vision techniques, and wireless communication technologies, these solutions can be integrated in a modular and low-cost manner. This work introduces Integrated Physical Security System (IP2S), a multi-agent system capable of coordinating diverse Internet of Things (IoT) sensors and actuators for an efficient mitigation of multiple physical security events. The proposed system was tested in a live case study that combined fire and intrusion detection in an industrial shop floor environment with four different sectors, two surveillance cameras, and a firefighting robot. The experimental results demonstrate that the integration of several events in a single automated system can be advantageous for the security of smart buildings, reducing false alarms and delays.

## 1 Introduction

Physical security is the protection of property and personnel from physical events such as theft, vandalism, fires, floods. By overlooking one of these events and not investing in adequate mechanisms, an organization can suffer severe damage and financial loss [1]. A study [2] reported a 20% increase of physical security incidents in 2021, in comparison with the previous year, a number that is forecasted to grow in 2022. Regarding fire hazards, the World Fire Statistics Center recorded an average of three and a half million fires and forty thousand deaths per year, from 1993 to 2019 [3]. These reports highlight the need for better and more robust physical security.

Modern organizations are starting to pay attention not only to straightforward threats, such as fire hazards and flood risks, but also to more intricate issues regarding surveillance and unauthorized personnel. Conventional solutions produce a high number of false positives and are usually in place from a detection point of view, activating warning alarms and asking for confirmation and actions to the active personnel. These lack of capabilities to automatically confirming events and placing countermeasures can lead to the rapid escalation of the situation. Consequently, security systems are becoming a crucial part of smart buildings [4].

This work builds on recent advancements on low-cost IoT sensors and communication protocols and contributes to the real-time security of intelligent buildings by introducing Integrated Physical Security System (IP2S), a simple, yet scalable multi-agent system capable of integrating different detection and mitigation agents, such as robust conventional sensor systems, real-time video-based fire detection (VFD), active surveillance systems and autonomous fire extinguishing robots. The system was evaluated in a live case study using an industrial shop floor environment, with intelligent low-cost sensors, rotating cameras, and a mixed firefighting surveillance robot.

The present paper is organized into multiple sections. Section 2 provides a survey of previous work on physical security systems, focused on fire and intrusion detection. Section 3 describes the proposed system and the concepts it relies on. Section 4 presents the case study and an analysis of the experimental results. Finally, Section 5 addresses the main conclusions and future work.

## 2    State-of-the-art

In smart buildings, there are several security threats that need to be addressed. Even though fires are not common events, they can cause severe damage to an organization and endanger the lives of its personnel. Conventional fire alarm systems are commonly limited to indoor spaces, have a lot of false positives, they cannot provide real-time information to firefighters [5] and, most importantly, have a lot of latency between the start of the fire and its detection [6].

Research advancements in fire detection introduce VFD supported by computer vision methods [7] as a reliable solution for early fire detection in real-world scenarios, such as forest fires [8], smart buildings and factories [9]. This technique can immediately detect when smoke or flames occur in one of the camera views, eliminating the threshold delay that the traditional alarms suffer, reduce the false positive rates, and transmit real-time crucial information about the fire to firefighters [6].

The progress on fire detection technologies, low-cost sensors and advancements in communications technologies permits the advancements in physical security systems. For instance, [10] prototyped a VFD implementation with a low-cost wireless sensor network. Moreover, [11] introduced a fire detection system, based on an Arduino microcontroller and a Raspberry Pi microprocessor. Despite these developments, the current literature lacks the integration of VFD and conventional fire detection systems with automatic and more precise suppression actuators, such as fire extinguishing robots that can access close-quarters and therefore prevent risk to human firefighters.

Regarding surveillance, traditional approaches include the use of passive infrared sensors and balanced magnetic switches [12] to detect intrusions and raise alerts. Closed-circuit Television (CCTV) cameras are also used for video surveillance of physical spaces for monitoring purposes [13], [14]. Even though they can be good solutions for recording multiple events, these cameras record vast amounts of footage, which makes it extremely time-consuming or even impossible to verify all the data captured producing a high cost on human resources.

Latest advances in research automate the process of analyzing the footage resorting to computer vision methods. For instance, [15] proposed an intelligent CCTV surveillance application in the context of a house to keep children safe, which focuses on detection and identification techniques. [16] proposed an image-based intrusion detection system capable of detecting intruders past a virtual fence by tracking and them and performing motion recognition in the context of nuclear power plants. [17] built a smart surveillance monitoring system on top of a Raspberry Pi 4 and other low-cost devices, resorting to computer vision techniques to identify the intruder, by analyzing frames received from a 360-degree rotating camera.

Despite the recent developments on smart surveillance, the current solutions only alert human personnel, without incorporating actuators to prevent further damage. This issue can be addressed by IoT devices, which have been converging information technology and operational technology. The use of artificial intelligence has been improving the security and efficiency of IoT devices by making them more reliable and robust [18], [19] and enabling a more adequate use of the available resources [20]. Additionally, by sharing data and resources, false alarms and the delay before a measure is taken can be reduced, taking more adequate mitigation measures [21].

An integrated physical security system must account for the specificities of different threats, which can be achieved via multiple interacting agents. Therefore, the multi-agent paradigm has the potential to significantly improve the reliability of security systems [22], but the divergent procedures required for different threats causes their integration to face challenges of autonomy, interoperability, and verification [23].

Since existing security systems usually contain surveillance cameras, [24] proposed their use for VFD as well. The authors considered the trajectory of a fire into the surrounding environment to perform a more thorough analysis of multiple footages. Nonetheless, fire detection remained independent from surveillance, so camera movement could not be controlled to better focus on the spread of fire. [25] attempted further integration by combining regular fire and intrusion detection with sensors of energy consumption and environmental factors in a single system. Each aspect was assigned to an agent, which performed an individual diagnosis and then shared the results. However, the agents did not interact with each other in an intelligent manner, so they could not use the data of the other agents to improve their own analysis.

To benefit from the integration of multiple physical security aspects, a system must be able to automatically use the data of one to improve and complement another. This is the gap in the current literature addressed by the proposed system. To the best of our knowledge, no previous work has introduced a similar approach.

## 3 Proposed System

IP2S was developed to integrate multiple physical aspects of the security of smart buildings. A modular design was adopted to enable the detection of multiple simultaneous physical security events in a variety of environments, with the only requirements being the logical division of an environment into sectors and the use of smart sensors and actuators (see Fig. 1). The following subsections detail the architecture and the physical implementation of the system with low-cost sensors to address the fire and intrusion detection aspects.

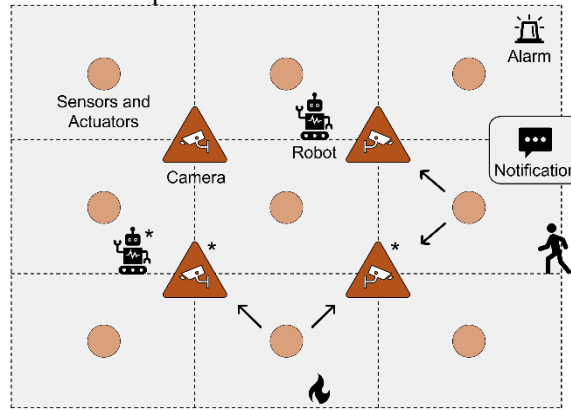

**Fig. 1.** Overview of the proposed system.

### 3.1 Multi-agent Architecture

The proposed system required the use of specialized agents that could perceive their environment and distribute different tasks spatially and temporally, where some agents could take actions immediately while others were required to wait for confirmation and feedback from their neighbors. To support multiple simultaneous physical security events, five distinct types of agents were created:

- **Sector Agent -** A hybrid agent that processes readings from a group of sensors and can take event-specific actions, such as triggering fire suppressors.
- **Camera Agent -** A collaborative agent that surveils the physical environment and analyzes the live feed to provide visual confirmation of ongoing events.
- **Robot Agent -** A mobile agent that can move throughout the physical environment to deploy event-specific mitigation measures in close-quarters.
- **Alarm Agent -** A reactive agent that places the system in an alarm state and activates adequate agents when an event is confirmed.
- **Notification Agent -** An interface agent that delivers real-time information about the system and its agents to authorized personnel.

The interactions between these agents ultimately lead to an event being mitigated or to a false alarm being disregarded. This flow was structured in a highly scalable architecture with both horizontal and vertical layers, where the Sector agents competed against each other, while the Camera and Robot agents cooperated (see Fig. 2). To deploy fully autonomous agents capable of performing the required negotiation, the *Spade* platform was utilized on top of the Python programming language.

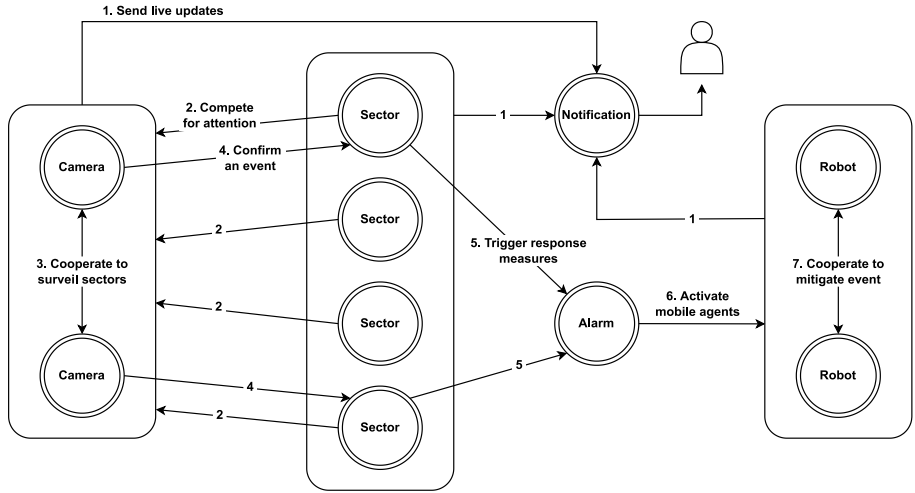

**Fig. 2.** Logical system architecture.

### 3.2 Physical Implementation

The sensors and actuators initially supported by IP2S had to be built in an efficient and scalable manner. ESP32 was the chosen board for the implementation, as it is a low-cost and low-power microcontroller. To bridge the cyber and physical spaces, the Message Queuing Telemetry Transport (MQTT) protocol was employed.

Each sector contains an ESP32 with a PIR motion detection sensor to detect movement, a DHT11 sensor to read temperature and humidity values, a push-up button for physical alarm activation and an LED, which represents the sector alarm. The cameras rely on their ability to rotate 360 degrees, so a step motor with a motor driver were used together with an ESP32 camera module. The microcontroller implementation uses a 2-core programming logic for better system performance. These cores enable a web server with the live surveillance feed, that is isolated from the MQTT communication and motor rotation tasks.

Finally, the robot is built on a 2-wheel drive chassis, with a motor in each wheel connected to a dual motor driver. It relies on three line-following sensors for navigation, which detect their black lone path by emitting infrared light. When requested, the robot can enter in surveillance or firefighting modes. In the first, the robot rapidly follows the line with the buzzer ringing and lights flashing until it reaches the specified sector. Once in that area, the robot slows down and performs a surveillance round, before returning to the parking station. In the latter, the robot starts searching for a fire when the specified sector is reached. It uses three infrared sensors to navigate itself around the fire, and when an adequate position is reached, a water pump and servo motor are activated to extinguish it.

Following the approach used for the cameras, both robot microprocessor cores are used. To efficiently run the line-following algorithm, interruptions must be kept to the minimum, which was achieved by having one system core completely dedicated to this task. At the other core, the Wi-Fi connection is maintained, the MQTT messages are handled, and lights and buzzers activated according to the event.

## 4     Case Study

A live case study was conducted to analyze the behavior of the proposed system. An industrial shop floor topology was used to represent a real-world dynamic environment where complex physical security events can occur. Two scenarios were analyzed: False Alarm and Dual Alarm. The first examines how the system reacts and confirms a false alarm triggered in one of the sectors, whereas the latter evaluates the system capabilities of reacting to two simultaneous events. The following subsections describe the environment and present an analysis of the obtained results.

### 4.1     Shop Floor Environment

The adopted topology divided the shop floor environment into four different sectors, each equipped with proximity sensors. To monitor the four sectors, a 360º camera was placed in the middle of them and an additional 180º camera was placed on the side, only capable of monitoring Sector2 and Sector3. The algorithms used by the camera were pretrained You Only Look Once version 3 (YOLOv3) models [26], provided by the *ImageAI* library. A robot was placed on a tape circuit throughout the sectors, with its parking station placed between Sector1 and Sector4 (see Fig. 3).

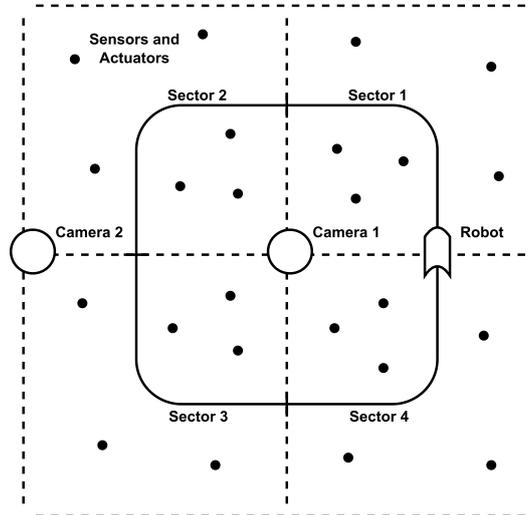

**Fig. 3.** Case study topology.

## 4.2 Results and Discussion

To perform a thorough analysis of the proposed system's performance, the behavior of each agent was recorded during the live experimentation. For each scenario, a single timeline was constructed with the interactions that occurred at each timestep, as well as the physical actions they resulted in.

In the False Alarm scenario, the cameras started following the regular activity throughout the shop floor by alternating between the different sectors where movement was detected. Then, when Sector3 recorded an abnormally high temperature and low humidity, in comparison with previous sensor readings, it messaged the cameras to request their confirmation that a fire is starting. The first to act was Camera2, which signaled that it was going to surveil that sector. Nonetheless, it was still the top priority for Camera1, so both cameras examined Sector3 without detecting any fire. As the temperature dropped in the subsequent readings, this false alarm was disregarded. Table 1 summarizes the key steps of this scenario.

**Table 1.** Timeline of the False Alarm scenario.

| # | Agent | Trigger | Action |
|---|-------|---------|--------|
| 1 | Sector2 | Sensors detected movement | Request attention from nearby cameras |
| 2 | Sector3 | Sensors read abnormal values | Request attention from nearby cameras |
| 3 | Sector4 | Sensors detected movement | Request attention from nearby cameras |
| 4 | Camera2 | Receive attention requests | Update its priority list, signalling that it is moving to Sector3 |
| 5 | Camera1 | Receive attention requests and Camera2 signal | Update its priority list, signalling that it is moving to Sector3 |
| 6 | Sector2 | Sensors detected movement | Request attention from nearby cameras |
| 7 | Sector4 | Sensors detected movement | Request attention from nearby cameras |
| 8 | Camera2 | Receive attention requests | Update its priority list, signalling that it is moving to Sector3 |
| 9 | Camera1 | Receive attention requests and signal | Update its priority list, signalling that it is moving to Sector3 |

In the Dual Alarm scenario, the cameras also started following regular activity. Then, Sector3 was locked down and the system received the signal to start considering any movement as intruder activity. After some time, Sector3 detected movement again. However, at the same timestep, Sector2 recorded abnormal values. Despite both attention requests having a high priority, the cooperation mechanism of the cameras enabled them to surveil both sectors. Since Camera1 was the first to act this time, its top priority was the sector with the possible fire, Sector2, which then led Camera2 to pick Sector3. The fire detection algorithm of Camera1 detected flames and a confirmation was sent to Sector2, which triggered the fire suppressors and started the process of activating the Robot. After performing close-quarters firefighting in Sector2, it advanced into Sector3 to help check for intruders. Table 2 summarizes these key steps.

**Table 2.** Timeline of the Dual Alarm scenario.

| # | Agent | Trigger | Action |
|---|---|---|---|
| 1 | Sector3 | Sensors detected movement | Request attention from nearby cameras |
| 2 | Sector3 | Lockdown signal received | Change its internal reaction state |
| 3 | Sector3 | Sensors detected movement | Request attention from nearby cameras and trigger intruder alarm |
| 4 | Sector2 | Sensors read abnormal values | Request attention from nearby cameras |
| 5 | Alarm | Receive Sector3 trigger | Activate intruder measures and Robot |
| 6 | Robot | Receive activation | Start moving to surveil Sector3 |
| 7 | Camera1 | Receive attention requests | Update its priority list, signalling that it is moving to Sector2 |
| 8 | Camera2 | Receive attention requests and Camera1 signal | Update its priority list, signalling that it is moving to Sector3 |
| 9 | Camera1 | Algorithm detected flames | Send fire confirmation to Sector2 |
| 10 | Sector2 | Receive fire confirmation | Trigger fire alarm |
| 11 | Alarm | Receive Sector2 trigger | Activate firefighting measures and a new request for Robot |
| 12 | Robot | Receive activation | Update course to first perform firefighting in Sector2 |

The obtained timelines evidence that the system not only detects complex emergency events, but that it can also confirm them and activate the most appropriate mitigation measures, reducing safety risks and diminishing associated costs.

## 5   Conclusions

This work addressed the physical security problematic by introducing IP2S, a multi-agent system that coordinates and integrates a low-cost IoT infrastructure for the automatic and efficient detection and mitigation of multiple threats. A live case study was conducted to evaluate the capabilities of the proposed system, using a simulated industrial shop floor environment, where intelligent sector proximity sensors, rotating cameras and actuator robot were deployed.

The experimental results obtained in the case study demonstrate that the system can accurately detect and confirm threats using proximity sensors and intelligent algorithms, which can significantly reduce the number of costly false alarms. Furthermore, IP2S can automatically and reliably deploy event-specific mitigation measures, which also reduces the delay before a fire is suppressed or an intruder is stopped from causing further damage to a smart building.

The integration of cyber and physical environments is no easy task, as there are many posing challenges that must be considered, such as the lack of accurate data, the dynamic nature of the environment, and their limited sphere of influence. In the future, with IP2S as the basis, more thorough solutions can be developed on top of its multi-agent paradigm, and more complex physical security threats can be addressed while maintaining a low-cost and modular approach.

**Acknowledgments.** The present work was partially supported by the Norte Portugal Regional Operational Programme (NORTE 2020), under the PORTUGAL 2020 Partnership Agreement, through the European Regional Development Fund (ERDF), within project "Cybers SeC IP" (NORTE-01-0145-FEDER-000044). This work has also received funding from UIDB/00760/2020. The case study was conducted on the facilities made available by the Polytechnic of Porto.

# References

[1] M. Jouini, L. B. A. Rabai, and A. Ben Aissa, "Classification of Security Threats in Information Systems," *Procedia Comput. Sci.*, vol. 32, pp. 489–496, Jan. 2014, doi: 10.1016/J.PROCS.2014.05.452.

[2] Pro-Vigil, "The State of Physical Security Entering 2022," 2022. https://pro-vigil.com/blog/physical-security-incidents-climbing-as-we-enter-2022/.

[3] N. N. Bushlinksy, M. Ahrens, S. V. Sokolov, and P. Wagner, "World Fire Statistics," 2021. https://ctif.org/world-fire-statistics.

[4] M. binti Mohamad Noor and W. H. Hassan, "Current research on Internet of Things (IoT) security: A survey," *Comput. Networks*, vol. 148, pp. 283–294, Jan. 2019, doi: 10.1016/J.COMNET.2018.11.025.

[5] D. Berry, J. Tate, A. Mclaughlin, S. Trew, A. Baxter, and & Atkinson, "FireGrid: Integrated emergency response and fire safety engineering for the future built environment."

[6] A. E. Çetin *et al.*, "Video fire detection – Review," *Digit. Signal Process.*, vol. 23, no. 6, pp. 1827–1843, Dec. 2013, doi: 10.1016/J.DSP.2013.07.003.

[7] A. Gaur, A. Singh, A. Kumar, A. Kumar, and K. Kapoor, "Video Flame and Smoke Based Fire Detection Algorithms: A Literature Review," *Fire Technol.*, vol. 56, no. 5, pp. 1943–1980, Sep. 2020, doi: 10.1007/S10694-020-00986-Y/TABLES/2.

[8] P. Barmpoutis, P. Papaioannou, K. Dimitropoulos, and N. Grammalidis, "A Review on Early Forest Fire Detection Systems Using Optical Remote Sensing," *Sensors 2020, Vol. 20, Page 6442*, vol. 20, no. 22, p. 6442, Nov. 2020, doi: 10.3390/S20226442.

[9] J. Seebamrungsat, S. Praising, and P. Riyamongkol, "Fire detection in the buildings using image processing," *Proc. 2014 3rd ICT Int. Sr. Proj. Conf. ICT-ISPC 2014*, pp. 95–98, Oct. 2014, doi: 10.1109/ICT-ISPC.2014.6923226.

[10] K. Kanwal, A. Liaquat, M. Mughal, A. R. Abbasi, and M. Aamir, "Towards Development of a Low Cost Early Fire Detection System Using Wireless Sensor Network and Machine Vision," *Wirel. Pers. Commun.*, vol. 95, no. 2, pp. 475–489, Jul. 2017, doi: 10.1007/S11277-016-3904-6/FIGURES/10.

[11] A. Imteaj, T. Rahman, M. K. Hossain, M. S. Alam, and S. A. Rahat, "An IoT based Fire Alarming and Authentication System for Workhouse using Raspberry Pi 3," *ECCE 2017 - Int. Conf. Electr. Comput. Commun. Eng.*, pp. 899–904, Apr. 2017, doi: 10.1109/ECACE.2017.7913031.

[12]   S. Smith, J. Ellis, and R. Abrams, "Central alarm stations and dispatch operations," *Prof. Prot. Off.*, pp. 89–103, 2010, doi: 10.1016/B978-1-85617-746-7.00008-0.

[13]   R. Prakash and P. Chithaluru, "Active Security by Implementing Intrusion Detection and Facial Recognition," *Lect. Notes Electr. Eng.*, vol. 692, pp. 1–7, 2021, doi: 10.1007/978-981-15-7486-3_1.

[14]   D. Lohani, C. Crispim-Junior, Q. Barthélemy, S. Bertrand, L. Robinault, and L. T. Rodet, "Perimeter Intrusion Detection by Video Surveillance: A Survey," *Sensors 2022, Vol. 22, Page 3601*, vol. 22, no. 9, p. 3601, May 2022, doi: 10.3390/S22093601.

[15]   K. Kanthaseelan *et al.*, "CCTV Intelligent Surveillance on Intruder Detection," *CCTV Intell. Surveill. Intruder Detect. Artic. Int. J. Comput. Appl.*, vol. 174, no. 14, pp. 975–8887, 2021, doi: 10.5120/ijca2021921035.

[16]   S. H. Kim, S. C. Lim, and D. Y. Kim, "Intelligent intrusion detection system featuring a virtual fence, active intruder detection, classification, tracking, and action recognition," *Ann. Nucl. Energy*, vol. 112, pp. 845–855, Feb. 2018, doi: 10.1016/J.ANUCENE.2017.11.026.

[17]   S. Choudhary, A. D. R, H. Likith Sai Varma, and L. S. Kumar, "Smart Surveillance Monitoring System using Machine Learning and Raspberry Pi," *International Research Journal of Modernization in Engineering*, vol. 585. .

[18]   F. Hussain, R. Hussain, S. A. Hassan, and E. Hossain, "Machine Learning in IoT Security: Current Solutions and Future Challenges," *IEEE Commun. Surv. Tutorials*, vol. 22, no. 3, 2020, doi: 10.1109/COMST.2020.2986444.

[19]   J. Vitorino, R. Andrade, I. Praça, O. Sousa, and E. Maia, "A Comparative Analysis of Machine Learning Techniques for IoT Intrusion Detection," in *Foundations and Practice of Security*, 2022, pp. 191–207, doi: 10.1007/978-3-031-08147-7_13.

[20]   H. F. Rashvand, K. Salah, J. M. A. Calero, and L. Harn, "Distributed security for multi-agent systems-Review and applications," *IET Inf. Secur.*, vol. 4, no. 4, pp. 188–201, Dec. 2010, doi: 10.1049/IET-IFS.2010.0041/CITE/REFWORKS.

[21]   Z. Liu and A. Kim, "Development of fire detection systems in the intelligent building," 2014.

[22]   J. Xie and C.-C. Liu, "Multi-agent systems and their applications," *J. Int. Counc. Electr. Eng.*, vol. 7, no. 1, pp. 188–197, 2017, doi: 10.1080/22348972.2017.1348890.

[23]   M. Pěchouček and V. Mařík, "Industrial deployment of multi-agent technologies: review and selected case studies," *Auton. Agents Multi-Agent Syst. 2008 173*, vol. 17, no. 3, pp. 397–431, May 2008, doi: 10.1007/S10458-008-9050-0.

[24]   N. M. Dung and S. Ro, "Algorithm for fire detection using a camera surveillance system," *ACM Int. Conf. Proceeding Ser.*, pp. 38–42, Feb. 2018, doi: 10.1145/3191442.3191450.

[25]   R. C. Luo, S. Y. Lin, and K. L. Su, "A multiagent multisensor based security system for intelligent building," *IEEE Int. Conf. Multisens. Fusion Integr. Intell. Syst.*, vol. 2003-Janua, pp. 311–316, 2003, doi: 10.1109/MFI-2003.2003.1232676.

[26]   J. Redmon and A. Farhadi, "YOLOv3: An Incremental Improvement," Apr. 2018, doi: 10.48550/arxiv.1804.02767.